\documentclass[reprint,superscriptaddress,amsmath,amssymb,aps,pre,nofootinbib,floatfix,longbibliography]{revtex4-2}
\usepackage{amsmath,amssymb,bm}
\usepackage[final]{graphicx}
\usepackage{booktabs}
\usepackage{placeins}
\usepackage[bookmarks=false,colorlinks=true,linkcolor=blue,citecolor=blue,urlcolor=blue]{hyperref}

\begin{document}
\title{Thermoresponsive self-oscillating hydrogel: spatially lumped model, relaxation dynamics and mechanical output}
\author{Yunjie Wang}
\affiliation{College of Chemical Engineering, China University of Mining and Technology, Xuzhou, 221116 Jiangsu, China}
\author{Ling Yuan}
\affiliation{College of Chemical Engineering, China University of Mining and Technology, Xuzhou, 221116 Jiangsu, China}
\author{Yang Li}
\affiliation{College of Chemical Engineering, China University of Mining and Technology, Xuzhou, 221116 Jiangsu, China}
\author{Lin Ren}
\thanks{Contact author: rlin1981@163.com}
\affiliation{College of Chemistry and Materials Engineering, Wenzhou University, Wenzhou, 325035 Zhejiang, China}
\author{Yuhan Wang}
\affiliation{College of Chemical Engineering, China University of Mining and Technology, Xuzhou, 221116 Jiangsu, China}
\author{Xiaol Lv}
\affiliation{College of Chemical Engineering, China University of Mining and Technology, Xuzhou, 221116 Jiangsu, China}
\author{Irving R Epstein}
\thanks{Contact author: epstein@brandeis.edu}
\affiliation{Department of Chemistry and Volen Center for Complex Systems, Brandeis University, Waltham, 02454-9110. Massachusetts, United States}
\author{Qingyu Gao}
\thanks{Contact author: Gaoqy@cumt.edu.cn}
\affiliation{College of Chemical Engineering, China University of Mining and Technology, Xuzhou, 221116 Jiangsu, China}

\begin{abstract}
Thermoresponsive gels with a lower critical solution temperature (LCST) can undergo self-oscillation driven by a thermal flux through repeated cycles of positive feedback, driven by heating-induced collapse, and negative feedback, characterized by deformation recovery due to collapse-induced suppression of heating and heat dissipation to the environment.
In order to explore the detailed bifurcation structure of a thermosensitive self-oscillating gel and understand its mechanical output of peristaltic motion and net displacement, we develop a minimal spatially lumped two-variable phenomenological model whose equilibrium parameters are constrained by poly(\textit{N}-isopropylacrylamide) (PNIPAAm) swelling data.
A supercritical Hopf bifurcation creates stable small-amplitude oscillations near a fold of the swollen manifold, and canard-mediated growth transforms this branch into relaxation oscillations.
Singular analysis relates the amplitude to fast jumps and the period to drift along slow manifolds, while continuation separates the Hopf boundary from the maximal-canard locus.
For the representative maximum-power relaxation cycle considered here, the loaded model predicts greater work than a harmonic volume trajectory with the same stroke, period, and viscous load, with output concentrated in the fast collapse and re-swelling transitions.
The net displacement during peristaltic locomotion of the thermoresponsive self-oscillating gel results from the work difference between collapse and re-swelling.
These results provide a theoretical basis for designing autonomous thermoresponsive self-oscillating gel actuators whose oscillatory waveform and power output can be controlled through thermal feedback and mechanical loading.
\end{abstract}

\maketitle

\section{Introduction}
Thermoresponsive gels provide a simple setting in which nonlinear mechanics, volume phase transition, and energy exchange can combine to produce autonomous motion~\cite{tanaka1978collapse,shibayama1993volume,hirokawa1984volume,zhang2017thermoresponsive,lee2020hydrogel,zhang2023thermoresponsive,zhao2021somatosensory}. In thermoresponsive polymer networks with thermal hysteresis, temperature variations can drive large swelling--collapse transitions, while cooling restores the swollen state~\cite{schild1992pnipam,afroze2000phase}. When a gel is subject to sustained internal heating and simultaneous environmental cooling, and when the collapsed state suppresses the heat input, these ingredients close a negative thermo-swelling feedback loop capable of generating self-sustained oscillations~\cite{he2012synthetic,zhao2019soft,zhang2022feedback,zhao2023sunlight,chung2025selfregulating}. In this class of systems, the oscillatory mechanism does not require an intrinsically oscillatory chemical reaction; rather, it arises from the coupling among temperature-dependent swelling, hysteretic volume change, feedback-controlled heat flux and heat dissipation to the environment.

Autonomous oscillations in responsive soft materials have been demonstrated and modeled in several contexts, including chemically driven oscillatory hydrogels and thermally driven oscillatory soft matter~\cite{he2012synthetic,zhang2022feedback,yoshida1996self,yoshida2010self}. For thermo-swelling oscillators, up to now, the theoretical picture remains incomplete. Existing reduced models for thermal or photothermal soft oscillators often emphasize structural vibration, heat transport, optical gating, or empirical feedback laws, without building an analytical bifurcation theory around the local nonlinear swelling mechanism of a thermoresponsive self-oscillating gel~\cite{he2012synthetic,zhao2019soft,xuan2022photodriven}. Normal-form and bifurcation analyses have been applied to photosensitive Belousov--Zhabotinsky gels and nanocatalyzed reactions to characterize their nonlinear dynamics, oscillation amplitude and frequency, and energy and power densities~\cite{rajput2024chaos,rajput2023energy}. These chemical oscillators are powered by reaction kinetics rather than the heat flux and mechanical deformation feedback considered here. Conversely, spatially resolved models based on nonlinear gel swelling mechanics often retain deformation and infinite dimensional space terms such as stress gradients, diffusion and fluid flow~\cite{he2012synthetic,xuan2022photodriven,onuki1993theory,barriere2003kinetics,doi2009gel,zhang2009finite}. Such models are valuable for mechanistic simulations, but their high-dimensional structure and deformation dynamics make bifurcation analysis challenging and often fail to make explicit the nature of the positive and negative feedback mechanisms.

Spatially lumped mathematical models can provide a bridge between an infinite-dimensional deformation model and the onset manifold and evolution of Hopf bifurcation and canard transition. The deformation subsystem already possesses a folded equilibrium manifold~\cite{doi2009gel}, while the thermal balance supplies a slow feedback variable. Hopf bifurcation and manifold analysis can therefore identify oscillation onset and explain its amplitude growth and canard transition. An analogous Hopf--canard--relaxation sequence occurs in a reduced two-dimensional slow--fast Belousov--Zhabotinsky model~\cite{xu2023dynamics}, although its folded geometry is generated by reaction kinetics. These are fast--slow and canard phenomena, for which local normal form analysis, singular perturbation theory, and periodic-orbit continuation provide important, complementary insights~\cite{guckenheimer1983nonlinear,kuznetsov2004elements,kuehn2015multiple}.

\begingroup\clubpenalty=10000 In this paper, we study self-oscillations in a heat flux-driven thermoresponsive gel using a minimal spatially lumped two-variable phenomenological model. The model couples overdamped swelling kinetics to a lumped heat balance. The nonlinear swelling response arises from the interplay between network elasticity and temperature-dependent polymer--solvent mixing. A collapse-suppressed heating gate and heat dissipation close the thermo-swelling feedback loop. We first derive the fast--slow structure and show how a stable small-amplitude branch born at a supercritical Hopf point undergoes canard-mediated growth into a relaxation cycle. We then augment the unloaded oscillator with a work-conjugate external load, allowing cycle work, mean output power, optimal load, and loss of oscillation under load to be calculated. This final step connects the bifurcation theory to waveform-selectable thermal flux-driven self-oscillating soft matter~\cite{he2012synthetic,zhang2022feedback}.\par\endgroup

\section{Model}\label{sec:model}

We model an internally heated PNIPAAm gel that exchanges solvent and heat with an ambient bath at temperature $T_0$. For an unloaded specimen, assumed to swell homogeneously and isotropically, the state variables are the spatially averaged polymer volume fraction $\phi(t)\in(0,1)$ and the temperature $T(t)$. Neglecting internal profiles and detailed boundary transport, we write the spatially lumped dynamics as
\begin{align}
\frac{d\phi}{dt} &= \frac{1}{\tau_\phi}\big[\Pi_{\mathrm{el}}(\phi)-\pi_{\mathrm{osm}}(\phi,T)\big], \label{eq:0d_final_phi}\\
\frac{dT}{dt} &= \gamma f(\phi)-k(T-T_0). \label{eq:0d_final_T}
\end{align}
Equation~\eqref{eq:0d_final_phi} describes a phenomenological overdamped relaxation of the collective swelling coordinate $\phi$. This formulation is motivated by two-fluid descriptions of gel dynamics, in which mechanical driving forces are balanced by network--solvent friction~\cite{barriere2003kinetics}. The driving term is the difference between the dimensionless pressure contributions $\Pi_{\mathrm{el}}$ from network elasticity and $\pi_{\mathrm{osm}}$ from polymer--solvent mixing. At fixed $T$, free-swelling equilibria satisfy $\Pi_{\mathrm{el}}=\pi_{\mathrm{osm}}$. The swelling time scale $\tau_\phi$ incorporates the effects of collective mobility, specimen geometry, and boundary solvent transport~\cite{tanaka1979kinetics}.

The temperature $T$ is expressed in degrees Celsius, and the corresponding absolute temperature satisfies $T_K/\mathrm{K}=T/{}^\circ\mathrm{C}+273.15$.

For homogeneous isotropic swelling, the network-elastic contribution takes the form~\cite{he2012synthetic},
\begin{equation}
\Pi_{\mathrm{el}}(\phi) = c_0 v_0\!\left[\left(\frac{\phi}{\phi_0}\right)^{1/3}-\frac{\phi}{2\phi_0}\right],
\label{eq:Pi_el}
\end{equation}
where $c_0$ is the number density of elastically active strands, $v_0$ is the solvent molecular volume, and $\phi_0$ is the polymer volume fraction in the reference configuration. Differentiating the reference-volume mixing free energy while retaining the concentration dependence of the interaction parameter gives
\begin{equation}
\begin{aligned}
\pi_{\mathrm{osm}}(\phi,T)
&= -\!\left[\phi+\ln(1-\phi)+\chi(\phi,T)\phi^2\right. \\
&\quad\left.-\phi^2(1-\phi)\frac{\partial\chi}{\partial\phi}\right].
\end{aligned}
\label{eq:pi_osm}
\end{equation}
Here $\chi(\phi,T)$ is the PNIPAAm--water Flory--Huggins interaction parameter. We use the composition-dependent parameterization~\cite{afroze2000phase,cai2011mechanics}
\begin{equation}
\chi(\phi,T)=\chi_0(T_K)+\chi_1(T_K)\phi,
\label{eq:chi}
\end{equation}
where $\chi_0(T_K)=A_0+B_0T_K$ and $\chi_1(T_K)=A_1+B_1T_K$, with coefficients specified below. Thus $\partial\chi/\partial\phi=\chi_1(T_K)$.

Equation~\eqref{eq:0d_final_T} is a lumped heat balance, where $T_0$ is the constant temperature of the ambient bath, $\gamma$ is the effective plateau heating rate, $k$ is the cooling rate coefficient, and $1/k$ is the thermal relaxation time scale. Appendix~\ref{app:lumped_kinetics} discusses the continuum motivation and scope of the lumped swelling law and the heat-balance reduction.

The heating gate $f(\phi)\in(0,1]$ represents a generic negative thermo-swelling feedback: temperature-induced collapse suppresses further heat input~\cite{xuan2022photodriven,zhang2022feedback,zhao2023sunlight}. We use the shifted Hill-type form
\begin{equation}
f(\phi)=\frac{1}{1+\big[\rho\,\max(\phi-\phi_g,0)\big]^{n}}
\label{eq:gate}
\end{equation}
where $\phi_g$ marks the end of the unit-heating plateau, $\rho^{-1}$ sets the characteristic transition width in polymer volume fraction, and $n$ controls its sharpness. The half-height occurs at $\phi=\phi_g+\rho^{-1}$. Figure~\ref{fig:heating_gate} shows the gate for the parameter values used below. The regularity of $f(\phi)$ at $\phi=\phi_g$ is discussed in Appendix~\ref{app:lumped_kinetics}.

Unless otherwise stated, the material parameters used below are $\phi_0=0.048$, $c_0v_0=1.6425\times10^{-3}$, $v_0=3.0\times10^{-29}\,\mathrm{m^3}$, $A_0=-12.947$, $B_0=0.04496\,\mathrm{K}^{-1}$, $A_1=17.92$, and $B_1=-0.0569\,\mathrm{K}^{-1}$. The interaction coefficients are those in Eq.~\eqref{eq:chi}. $\phi_0$ is selected from the reported sample, and $c_0v_0$ is fitted to its stable swelling data~\cite{oh1998crosslink}, while the data of Suzuki \textit{et al.}~\cite{suzuki1999shrinking} provide a cross-material consistency check. Appendix~\ref{app:equilibrium_calibration} documents the data extraction, point selection, fitting objective, residuals, and conditional parameter uncertainty. The effective dynamic and feedback values are $T_0=25\,^\circ\mathrm{C}$, $\tau_\phi=0.1\,\mathrm{s}$, $\phi_g=0.139$, $\rho=30$, and $n=3$, while $k$ and $\gamma$ are varied as stated for each calculation.

\begin{figure*}
  \centering
  \includegraphics[width=\linewidth]{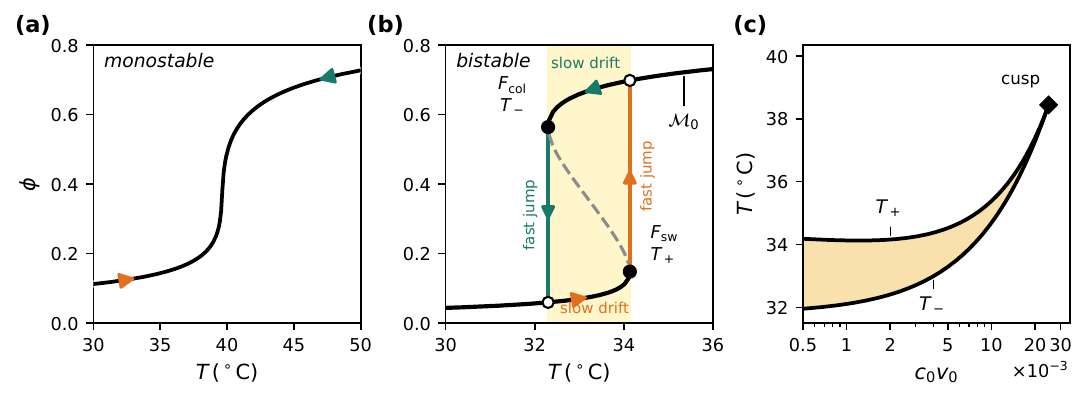}
  \caption{Steady swelling branches and the elasticity-controlled transition between hysteretic and single-valued thermal responses.
  In panels~(a) and (b), black solid and gray dashed curves denote stable and unstable equilibrium branches, while orange and green arrows denote heating and cooling.
  (a)~Hypothetical high-elasticity monostable control ($\phi_0=0.048$, $c_0v_0=3.0\times10^{-2}$); heating and cooling follow the same stable branch.
  (b)~Bistable lower-elasticity thermoresponsive gel. Literature-calibrated working point for the sample with cross-linker feed fraction $C=1.23\%$ ($\phi_0=0.048$, $c_0v_0=1.6425\times10^{-3}$); the fold points $F_{\rm col}$ and $F_{\rm sw}$ delimit $32.301$--$34.136\,^\circ$C. The solid and dashed equilibrium branches together form the critical manifold $\mathcal M_0$. Arrows trace slow drift and fast isothermal jumps of the singular relaxation orbit [Eq.~\eqref{eq:singular_limits}]; open circles mark the landing states.
  (c)~Phase diagram in the $(c_0v_0,T)$ plane with a logarithmic $c_0v_0$ axis. The lower and upper black curves are $T_-(c_0v_0)$ and $T_+(c_0v_0)$, respectively. The yellow region between them is bistable. The fold pair terminates at the cusp $(c_0v_0,T,\phi)=(2.4801\times10^{-2},38.442\,^\circ\mathrm{C},0.3681)$.}
  \label{fig:equilibrium_combined}
\end{figure*}

\section{Stability and Bifurcation Analysis}\label{sec:bifurcation}

\begingroup
\interlinepenalty=10000
We first identify the singular geometry set by the constitutive swelling
law, then determine how finite-rate dynamics generate and enlarge a stable
periodic orbit, and finally organize these states in the natural
two-parameter plane.\par
\endgroup

\subsection{Fast--slow structure and singular geometry}
\label{sec:fastslow_geometry}

\begin{figure*}[!t]
  \centering
  \includegraphics[width=\linewidth]{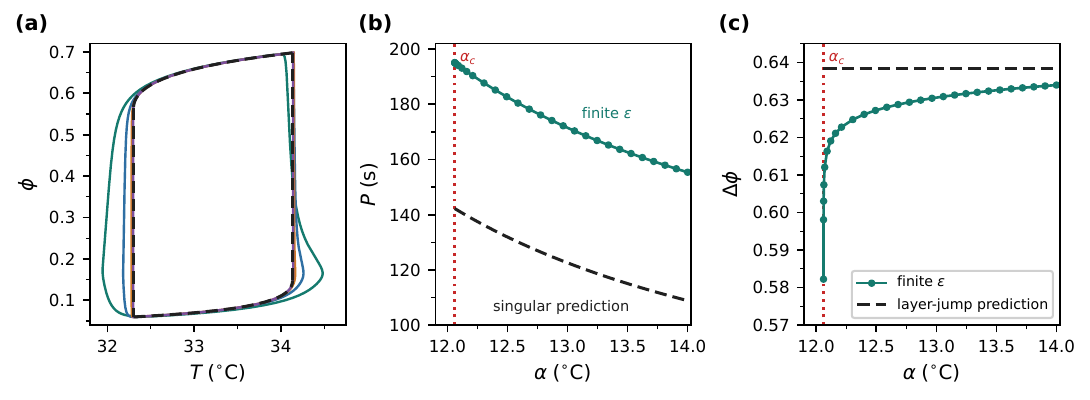}
  \caption{Singular-limit predictions and finite-rate convergence.
  (a)~Orbit convergence at $\alpha=14\,^\circ\mathrm C$.  The black dashed
  curve is the composite $\varepsilon=0$ relaxation orbit.  Teal, blue,
  orange, and purple curves denote
  $\varepsilon=5\times10^{-4}$, $10^{-4}$, $10^{-5}$, and $10^{-6}$,
  respectively, obtained by varying $\tau_\phi$ at fixed
  $k=5\times10^{-3}\,\mathrm{s}^{-1}$.
  (b)~Finite-$\varepsilon$ period and the singular slow-passage period prediction
  $P_0(\alpha)$ at $\varepsilon=5\times10^{-4}$; the latter is the reduced
  time accumulated on the two attracting branches.  The red dotted lines
  in panels~(b) and (c) mark the maximal-canard value $\alpha_c$ obtained in
  Sec.~\ref{sec:hopf_canard}.
  (c)~Corresponding oscillation amplitude and the layer-jump prediction
  $\Delta\phi_0$, defined by the separation of the two isothermal landing
  states.}
  \label{fig:singular_limit_prediction}
\end{figure*}

With the material and heating-gate parameters fixed, using the thermal relaxation
time scale $1/k$ as the time unit isolates two physically distinct controls,
\begin{equation}
\alpha\equiv\frac{\gamma}{k}\,,\qquad
\varepsilon\equiv\tau_\phi k.
\label{eq:natural_controls}
\end{equation}
where $\alpha$ is the steady temperature rise sustained by a fully open
heating gate and $\varepsilon$ is the ratio of the swelling time scale
$\tau_\phi$ to the thermal relaxation time scale $1/k$.  On defining the dimensionless time $s=kt$, the mechanical drive
$g(\phi,T)\equiv\Pi_{\mathrm{el}}(\phi)-\pi_{\mathrm{osm}}(\phi,T)$, and the thermal drive $h(\phi,T;\alpha)\equiv\alpha f(\phi)-(T-T_0)$,
eqs.~\eqref{eq:0d_final_phi} and \eqref{eq:0d_final_T} become
\begin{align}
\varepsilon\frac{d\phi}{ds}&=g(\phi,T),\label{eq:fastslow_phi}\\
\frac{dT}{ds}&=h(\phi,T;\alpha).\label{eq:fastslow_T}
\end{align}
In our system, $\varepsilon\ll1$, so that $\phi$ is fast and $T$ is slow.
The equilibria of the fast subsystem form the mechanical critical manifold
$\mathcal M_0=\bigl\{(\phi,T):g(\phi,T)=0\bigr\}$, which is the
free-swelling equilibrium curve of the mechanical subsystem. For fixed
material parameters, its geometry is independent of $\alpha$ and $\varepsilon$.
The parameter $\alpha$ sets the thermal nullcline and the available feedback
heating along $\mathcal M_0$, whereas $\varepsilon$ sets the degree of
mechanical--thermal time scale separation and the accuracy of the
quasistatic approximation on $\mathcal M_0$. Dimensional periods retain the
overall factor $1/k$.

Segments with $g_\phi\equiv\partial g/\partial\phi<0$ are attracting, whereas those with $g_\phi>0$ are
repelling. For the data-constrained parameter set, $\mathcal M_0$ is the
S-shaped curve shown in Fig.~\ref{fig:equilibrium_combined}(b), with
attracting swollen and collapsed branches separated by a repelling branch.
The fold points $F_{\rm sw}$ and $F_{\rm col}$, defined by $g=g_\phi=0$, occur at
$(\phi,T)=(0.1480,34.136\,^\circ\mathrm C)$ and
$(0.5636,32.301\,^\circ\mathrm C)$, respectively.
Figure~\ref{fig:equilibrium_combined} therefore establishes that the folded
critical manifold is a constitutive property of the equilibrium-calibrated
material rather than a consequence of the thermal feedback. The calibrated
value of $c_0v_0$ lies below its cusp value [Fig.~\ref{fig:equilibrium_combined}(c)],
so $\mathcal M_0$ has the two fold points required by the fast--slow mechanism
below; the fold calculation is given in Appendix~\ref{app:equilibrium_stability}.

The equilibria of the coupled system are intersections of
$\mathcal M_0$ with the thermal nullcline
$T=T_0+\alpha f(\phi)$.  At $\alpha_F=9.31615\,^\circ\mathrm C$,
the thermal nullcline passes through
$F_{\rm sw}$, so $h(F_{\rm sw};\alpha_F)=0$ at the same point where
$g_\phi(F_{\rm sw})=0$.  The resulting singular canard point is the
$\varepsilon\to0$ organizer of the Hopf and maximal-canard loci discussed
below.  The same critical-manifold geometry also determines the singular
relaxation orbit.  In the limit $\varepsilon\to0$, the fast time
$\theta=s/\varepsilon$ gives the layer and reduced
problems~\cite{kuehn2015multiple},
\begin{equation}
\begin{aligned}
\text{fast jump:}\quad&
\frac{d\phi}{d\theta}=g(\phi,T),&
\frac{dT}{d\theta}&=0,\\
\text{slow drift:}\quad&
0=g(\phi,T),&
\frac{dT}{ds}&=h(\phi,T;\alpha).
\end{aligned}
\label{eq:singular_limits}
\end{equation}

As indicated in Fig.~\ref{fig:equilibrium_combined}(b), slow drift follows the
attracting branches of $\mathcal M_0$, whereas fast isothermal jumps
terminate on the opposite attracting branch.  At the calibrated material
parameters, the fold temperatures in Fig.~\ref{fig:equilibrium_combined}(c)
are $T_-=32.301\,^\circ\mathrm C$ and $T_+=34.136\,^\circ\mathrm C$.
On each attracting branch, we may write $\phi=\phi_j(T)$ and
$H_j(T;\alpha)=h[\phi_j(T),T;\alpha]$, with $j=\mathrm{sw},\mathrm{col}$.
In the parameter range considered here, $\alpha>\alpha_F$ gives
$H_{\rm sw}>0$ and $H_{\rm col}<0$ between the folds.  The singular orbit
therefore heats along the swollen branch to $F_{\rm sw}$, collapses
isothermally at $T_+$, cools along the collapsed branch to
$F_{\rm col}$, and re-swells isothermally at $T_-$.

The stable landing states are
$\phi_{\rm sw}(T_-)=0.059566$ and $\phi_{\rm col}(T_+)=0.697817$,
marked by open circles in Fig.~\ref{fig:equilibrium_combined}(b).
In the singular limit $\varepsilon\to0$, the constitutive manifold therefore fixes the oscillation amplitude
$\Delta\phi_0=\phi_{\rm col}(T_+)-\phi_{\rm sw}(T_-)=0.638252$, whereas the
two slow passages determine the period
$P_0(\alpha)=k^{-1}\int_{T_-}^{T_+}
[1/H_{\rm sw}(T;\alpha)-1/H_{\rm col}(T;\alpha)]\,dT$.
At fixed $\alpha$, this gives $P_0\propto k^{-1}$; a fixed-$\gamma$ scan
does not preserve the scaling, because it changes both $\alpha$ and
$\varepsilon$.  Both predictions follow directly from the constitutive
manifold without fitting a finite-rate orbit.  For
$0<\varepsilon\ll1$, the regular portions of $\mathcal M_0$ persist as
nearby slow manifolds, but delayed departure through the fold layers
produces an $O(\varepsilon^{2/3})$ overshoot in $T$.
Figure~\ref{fig:singular_limit_prediction}(a) shows the
finite-rate cycle converging to the composite slow-drift/fast-jump orbit as
$\varepsilon$ decreases at fixed $\alpha=14\,^\circ\mathrm C$.  At
$\varepsilon=5\times10^{-4}$, the numerical period in
Fig.~\ref{fig:singular_limit_prediction}(b) remains above
$P_0(\alpha)$, whereas the amplitude in
Fig.~\ref{fig:singular_limit_prediction}(c) is already close to
$\Delta\phi_0$.  Additional residence time in the fold layers affects the period more
strongly than the jump-determined amplitude at the same $\varepsilon$.
Appendix~\ref{app:singular_convergence} verifies the expected fold-layer
convergence and the recovery of both singular predictions without changing
the constitutive parameters.

\begin{figure*}[!t]
  \centering
  \includegraphics[width=0.80\linewidth]{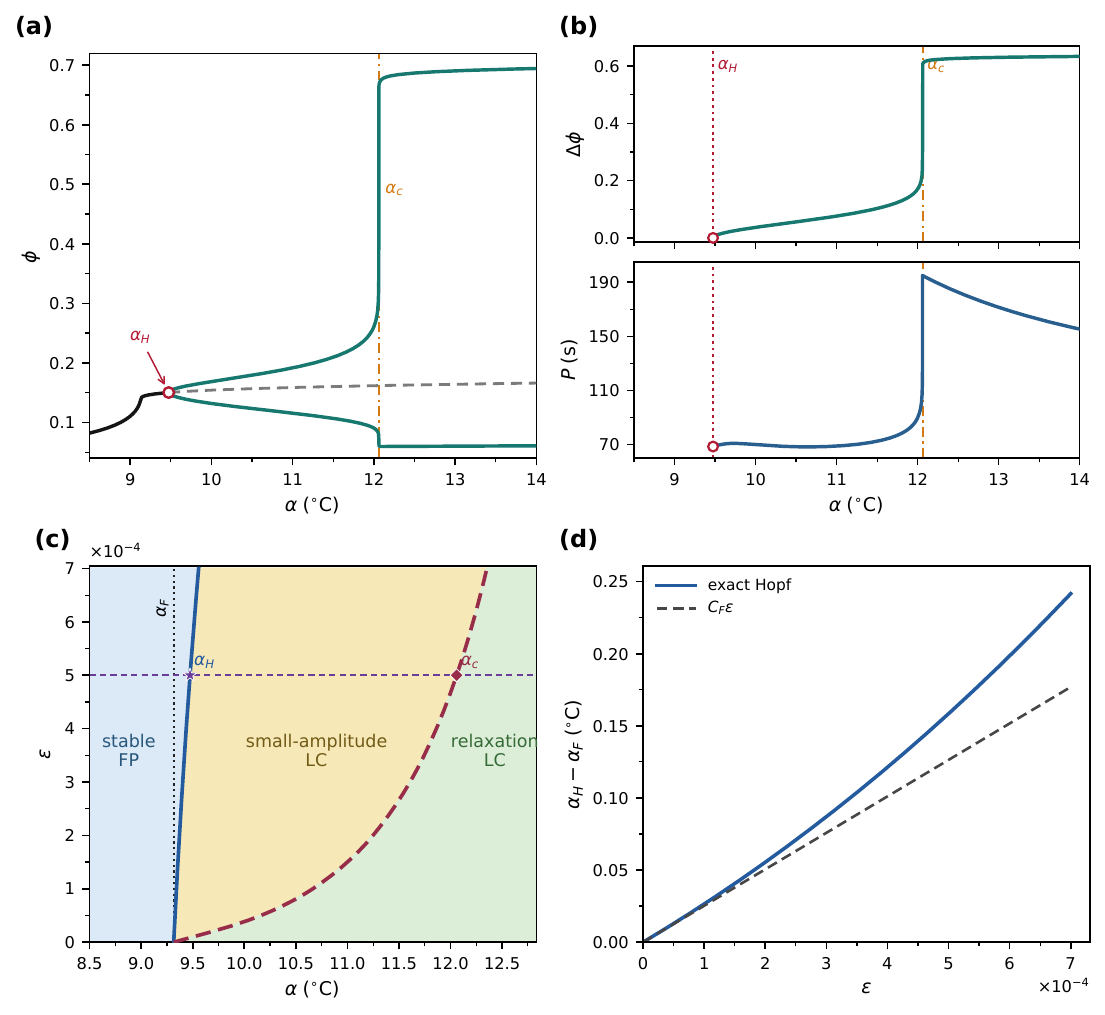}
  \caption{Hopf onset and canard organization of the oscillatory regimes.
  (a)~Bifurcation diagram. Stable (black) and unstable (gray dashed) fixed points and the extrema
  of the stable periodic branch (teal) at
  $\varepsilon=5\times10^{-4}$.  The open circle marks $\alpha_H$ and the
  orange dash-dotted line marks the maximal canard $\alpha_c$.
  (b)~Amplitude $\Delta\phi$ and period $P$ along the stable periodic branch in (a).
  (c)~Phase diagram in the $(\alpha,\varepsilon)$ plane.  The solid
  blue Hopf curve separates stable fixed points from stable limit cycles;
  the dashed burgundy maximal-canard curve organizes the crossover from
  small-amplitude to relaxation waveforms without a loss of cycle stability.
  The black dotted
  line is the singular-fold value $\alpha_F$, and the purple dashed line is
  the representative slice.
  (d)~Exact Hopf displacement from the fold compared with the singular
  prediction $\alpha_H-\alpha_F=C_F\varepsilon$.}
  \label{fig:hopf_canard_organization}
\end{figure*}

\subsection{Hopf onset and canard organization of relaxation oscillations}
\label{sec:hopf_canard}

At finite $\varepsilon$, the coupled fixed point loses stability only after
it has passed the swollen-side mechanical fold.  Linearizing
Eqs.~\eqref{eq:fastslow_phi} and \eqref{eq:fastslow_T} at $(\phi^\ast,T^\ast)$ gives the zero-trace and positive-determinant conditions for Hopf onset,
\begin{equation}
g_\phi^\ast=\varepsilon\quad\text{and}\quad
\alpha f'(\phi^\ast)\pi_T(\phi^\ast,T^\ast)>\varepsilon .
\label{eq:hopf_conditions}
\end{equation}
Here $\pi_T\equiv\partial\pi_{\mathrm{osm}}/\partial T$.
The first condition places the onset on the mechanically repelling branch, an
$O(\varepsilon)$ distance beyond $F_{\rm sw}$.  The second requires the
thermo-swelling negative feedback to be strong enough to produce a complex
eigenvalue pair rather than a saddle.  Expanding the equilibrium curve about
the fold gives
\begin{equation}
\alpha_H(\varepsilon)
=\alpha_F+C_F\varepsilon+O(\varepsilon^2),\qquad
C_F=252.688\,^\circ\mathrm C .
\label{eq:hopf_fold_prediction}
\end{equation}
For the representative slice $\varepsilon=5\times10^{-4}$, obtained with
$\tau_\phi=0.1\,\mathrm{s}$ and
$k=5\times10^{-3}\,\mathrm{s}^{-1}$, the full finite-$\varepsilon$
calculation gives $\alpha_H\simeq9.475\,^\circ\mathrm C$.  The negative first
Lyapunov coefficient identifies a supercritical Hopf bifurcation and hence
the birth of a stable small-amplitude limit cycle~\cite{kuznetsov2004elements}.
Numerical continuation follows this cycle along a single stable periodic
branch throughout the interval considered; no fold of periodic orbits is
detected.  The local derivation and normal form calculation are given in
Appendix~\ref{app:lyapunov}, and the continuation and stability checks are
described in Appendix~\ref{app:numerical_methods}.

\FloatBarrier

\begin{figure*}[!t]
  \centering
  \includegraphics[width=\linewidth]{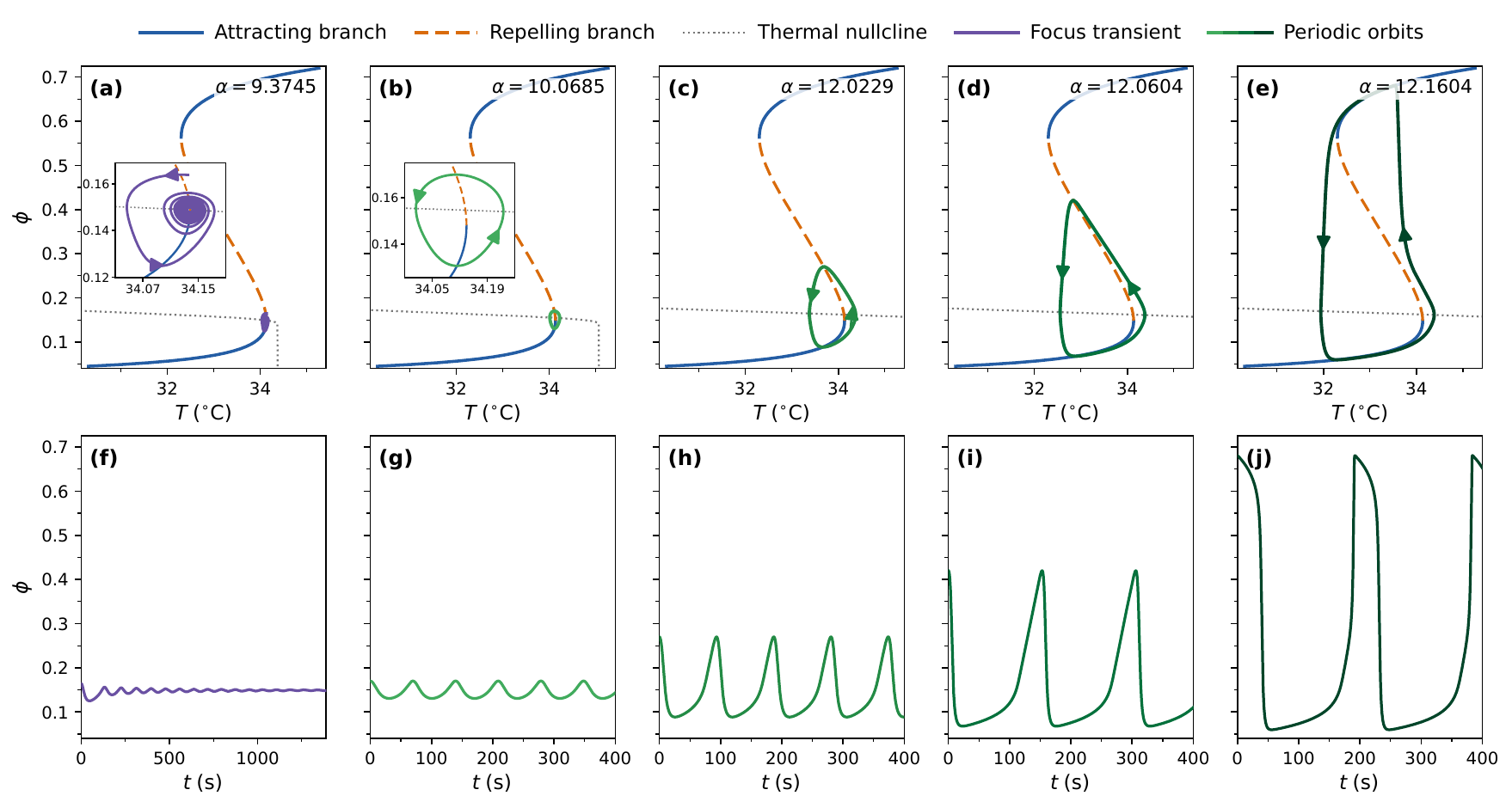}
  \caption{Phase portraits and time series at
  $\varepsilon=5\times10^{-4}$.  Panels (a)--(e) show the complete
  critical manifold, the thermal nullcline, and the trajectory; the
  insets in (a) and (b) magnify the contracting focus and the small periodic
  orbit.  Panels (f)--(j) show the corresponding $\phi(t)$ traces.
  From left to right the pairs represent (a,f) a stable focus before Hopf,
  (b,g) a nearly harmonic stable limit cycle, (c,h) a small
  relaxation-shaped cycle, (d,i) the maximal-canard orbit, and
  (e,j) a fully developed post-canard relaxation
  cycle.  The $\alpha$ value of each state is shown in its phase-plane panel.  Arrows indicate traversal direction.}
  \label{fig:canard_representative_states}
\end{figure*}

\begingroup
\widowpenalty=10000
Figures~\ref{fig:hopf_canard_organization}(a) and
\ref{fig:hopf_canard_organization}(b) show the stable periodic branch and the
variation of its amplitude and period with $\alpha$.  Its amplitude initially
grows smoothly from zero.  Near
$\alpha_c\simeq12.0604\,^\circ\mathrm C$, the orbit follows the repelling
slow manifold for an increasingly long interval, and the attracting and
repelling slow manifolds match to form the maximal canard.  The amplitude
then rises sharply and the period increases because of the additional
residence time near the repelling manifold and the fold bottleneck.  Once
the outer relaxation loop is established, the period decreases toward the
prediction based on the two slow passages described in Sec.~\ref{sec:fastslow_geometry}.  The
periodic branch remains smooth and stable at $\alpha_c$; the maximal canard
is therefore not a second bifurcation, but the geometric organizer of the
rapid amplitude growth~\cite{kuehn2015multiple,krupa2001relaxation}.  This distinction between
a stability boundary and a waveform organizer also structures the
two-parameter result in Fig.~\ref{fig:hopf_canard_organization}(c).  The Hopf curve
$\alpha=\alpha_H(\varepsilon)$ is the true bifurcation boundary
at which the fixed point loses stability.  The maximal-canard curve
$\alpha=\alpha_c(\varepsilon)$ instead organizes the waveform
change within the stable periodic regime.  The light blue domain in
Fig.~\ref{fig:hopf_canard_organization}(c) contains stable fixed points,
whereas the light yellow and light green domains contain stable limit cycles
with small-amplitude near-harmonic and relaxation waveforms, respectively.  Increasing
$\varepsilon$ shifts
both loci to larger $\alpha$, but $\alpha_c$ is more sensitive because weaker
timescale separation requires stronger feedback to assemble the outer
relaxation loop.  Both loci approach $\alpha_F$ as
$\varepsilon\rightarrow0$, while
Fig.~\ref{fig:hopf_canard_organization}(d) verifies the linear Hopf
displacement in Eq.~\eqref{eq:hopf_fold_prediction}.
Appendix~\ref{app:gate_robustness} verifies that parameter variations and
matched alternative gates preserve the progression from a supercritical Hopf
bifurcation through the maximal canard to relaxation oscillations, while the
thresholds and narrow-window Floquet stability remain closure dependent.\par
\endgroup

\FloatBarrier

The paired phase portraits and time traces in
Fig.~\ref{fig:canard_representative_states} make the corresponding geometric
conversion explicit: increasing residence near the repelling branch precedes
the opening of the outer relaxation loop, thereby connecting the local Hopf
instability to the large-stroke motion analyzed below.

\FloatBarrier

\section{Mechanical output under load}\label{sec:results}

\subsection{Work-conjugate load closure}\label{sec:load_model}

\begin{figure*}[!t]
  \centering
  \includegraphics[width=0.755\linewidth]{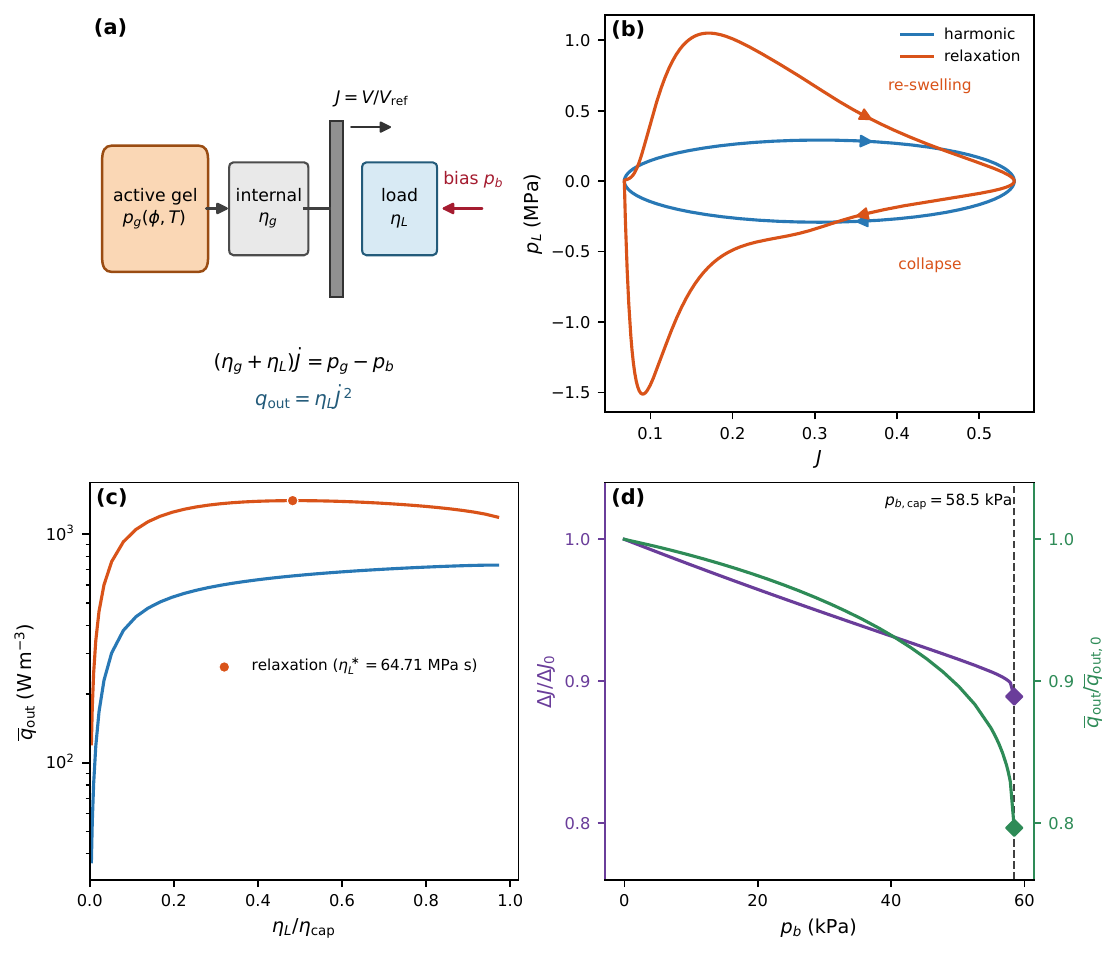}
  \caption{Mechanical output of the loaded thermoresponsive gel.
  (a)~Work-conjugate load model.  The active swelling pressure $p_g$ drives
  the volume ratio $J$ through the internal hydraulic viscosity $\eta_g$,
  the power-taking load $\eta_L$, and the opposing pressure $p_b$.
  (b)~Load-pressure--volume loops for the relaxation cycle (orange) and the
  prescribed harmonic reference (blue), with $p_L=\eta_L\dot J$ and matched
  volume extrema, period, and viscous load.  Their enclosed areas give the
  extracted work density per cycle in Eq.~\eqref{eq:mechanical_output_metrics}.  Arrows show the direction of traversal,
  with collapse on the lower branch and re-swelling on the upper branch.
  (c)~Mean output power density versus viscous load.  Both curves use the
  relaxation load capacity $\eta_{\mathrm{cap}}$ ($134.06\,\mathrm{MPa\,s}$) for normalization.  The
  harmonic reference (blue) is matched separately to the relaxation orbit
  (orange) at each load.  The orange circle marks the relaxation
  maximum-power point.
  (d)~Volume stroke and mean output power retained by the relaxation mode
  as the opposing pressure is increased at its maximum-power viscous load.
  Here $\Delta J$ is the volume ratio stroke over one cycle, and
  $\Delta J_0$ is its value at $p_b=0$ with the same viscous load.
  Diamonds and the dashed line mark the finite-amplitude pressure-load
  fold.
  The relaxation state uses $k=5\times10^{-3}\,\mathrm{s^{-1}}$ and
  $\gamma=0.0608018\,^\circ\mathrm C\,\mathrm{s^{-1}}$
  ($\varepsilon=5\times10^{-4}$, $\alpha=12.16036\,^\circ\mathrm C$).
  Pressure is dimensionalized using the molecular volume
  $v_0=3.0\times10^{-29}\,\mathrm{m^3}$ specified in Sec.~\ref{sec:model}.
  Panel~(b) uses $\eta_L=64.71\,\mathrm{MPa\,s}$ at $p_b=0$.
  Panel~(c) varies $\eta_L$ at $p_b=0$; panel~(d) varies $p_b$ at
  $\eta_L=64.71\,\mathrm{MPa\,s}$.}
  \label{fig:mechanical_output}
\end{figure*}

The unloaded model above determines the autonomous swelling waveform but does
not by itself define useful mechanical work.  A natural mechanical
coordinate is
\begin{equation}
J\equiv\frac{V}{V_{\mathrm{ref}}}=\frac{\phi_0}{\phi},
\label{eq:load_J}
\end{equation}
the ratio of current to reference gel volume, which is work conjugate to pressure.  With the dimensional osmotic-pressure
scale $p_T(T)=k_BT_K/v_0$, the pressure available to drive an external
device is
\begin{equation}
p_g(\phi,T)
=p_T(T)\big[\pi_{\mathrm{osm}}(\phi,T)-\Pi_{\mathrm{el}}(\phi)\big]
=-p_T(T)g(\phi,T).
\label{eq:gel_output_pressure}
\end{equation}
Rewriting the original swelling law in terms of $J$ gives
$\eta_g(\phi,T)\dot J=p_g$, where
$\eta_g=\tau_\phi p_T\phi^2/\phi_0$ is the state-dependent internal
hydraulic viscosity implied by the existing constant-$\tau_\phi$ closure.
This identity introduces no new constitutive assumptions.

An energy-extracting viscous load with dynamic viscosity coefficient $\eta_L$ (pressure per
unit deformation rate) and a constant opposing pressure $p_b$ then give
the loaded mechanical balance
\begin{equation}
\big[\eta_g(\phi,T)+\eta_L\big]\dot J=p_g(\phi,T)-p_b.
\label{eq:loaded_gel_balance}
\end{equation}
Figure~\ref{fig:mechanical_output}(a) illustrates this load model.
Equation~\eqref{eq:loaded_gel_balance}, together with the unchanged heat
balance in Eq.~\eqref{eq:0d_final_T}, closes the loaded two-state model and
reduces to Eq.~\eqref{eq:0d_final_phi} when
$\eta_L=p_b=0$.

The instantaneous extracted power per reference gel volume, the work
density per cycle, and the mean power density are
\begin{equation}
q_{\mathrm{out}}=\eta_L\dot J^2,\qquad
w_{\mathrm{out}}=\int_0^Pq_{\mathrm{out}}\,dt,\qquad
\overline q_{\mathrm{out}}=\frac{w_{\mathrm{out}}}{P}.
\label{eq:mechanical_output_metrics}
\end{equation}
With the load pressure $p_L=\eta_L\dot J$, the extracted work density per
cycle can also be written as $w_{\mathrm{out}}=\oint p_L\,dJ$.
The cycle also obeys the mechanical energy identity
\begin{equation}
\oint p_g\,dJ
=\int_0^P\eta_g\dot J^2\,dt+w_{\mathrm{out}},
\label{eq:mechanical_energy_identity}
\end{equation}
because $p_b\oint dJ=0$ for a closed orbit.  Thus a constant bias tests
load-bearing capacity but does no net work over a reciprocal volume cycle;
the dissipative element $\eta_L$ represents the power-taking load.

\subsection{Waveform-dependent power and load capacity}
\label{sec:waveform_output}

Figure~\ref{fig:mechanical_output} compares the mechanical output of the
loaded relaxation cycle with a prescribed harmonic volume trajectory.
The dimensional mechanical quantities reported
below are conditional predictions per reference gel volume, based on the
pressure scale and illustrative kinetic and thermal parameters specified in
the figure caption, rather
than absolute outputs calibrated for a particular specimen.
Periodic relaxation orbits were continued in $\eta_L$, and their stability was checked
from the nontrivial planar Floquet multiplier.  Appendix~
\ref{app:mechanical_output_numerics} gives details of continuation, fold location,
and numerical validation.

\begin{figure*}
  \centering
  \includegraphics[width=0.80\linewidth]{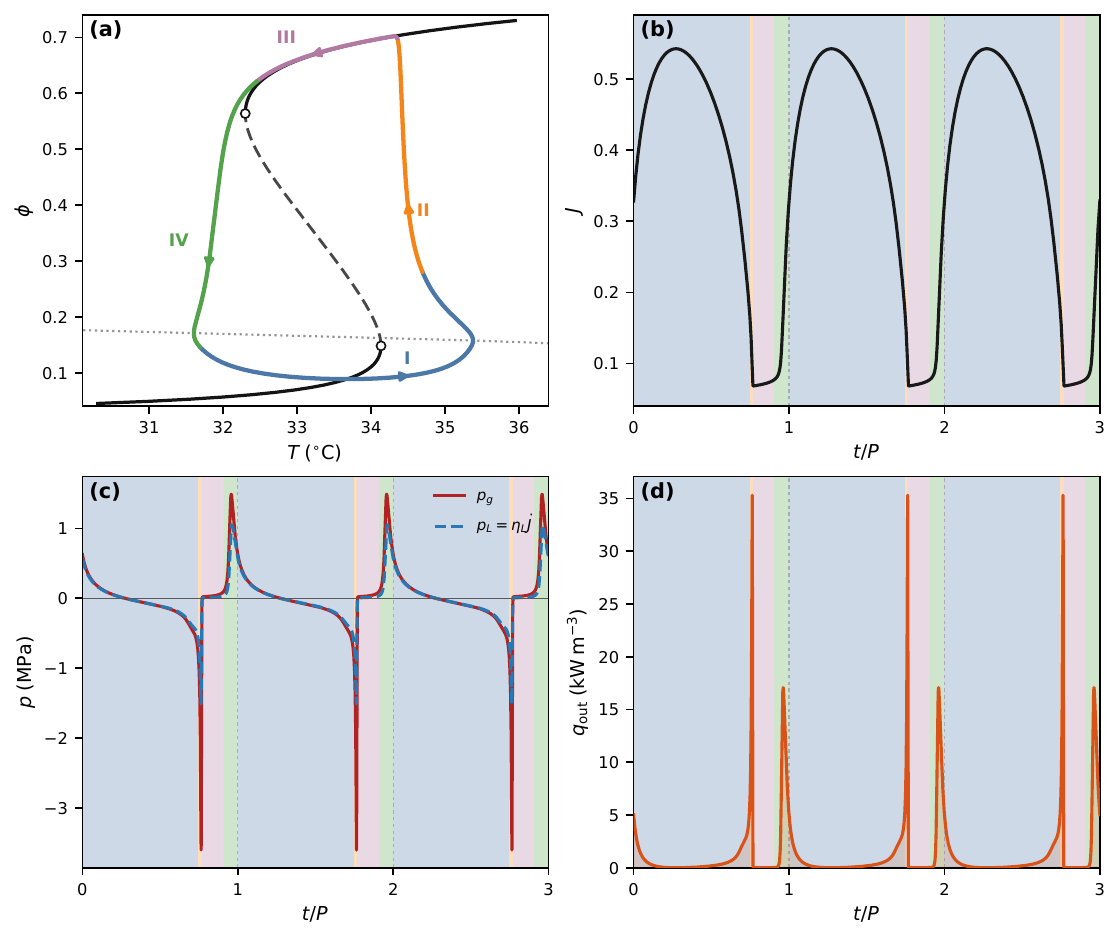}
  \caption{Stage-resolved mechanical output at the maximum-power relaxation
  point. Parameters are $\alpha=12.16036\,^\circ\mathrm C$,
  $\varepsilon=5\times10^{-4}$,
  $\eta_L=64.71\,\mathrm{MPa\,s}$, and $p_b=0$.  The orbit is rephased to
  start at Stage~I.  Colors identify I (slow swelling), II (fast collapse),
  III (slow collapse), and IV (fast re-swelling).
  (a)~The complete swelling critical manifold in the $(T,\phi)$ plane.
  Solid and dashed curves are its attracting and repelling parts,
  respectively, the gray dotted curve is the thermal nullcline, open
  circles mark the folds, and the colored arrows show the four stages of the loaded orbit.
  (b)~Volume ratio over three loaded cycles.
  (c)~Gel pressure and the pressure transmitted to the dissipative load,
  $p_L=\eta_L\dot J$.
  (d)~Instantaneous output-power density over the same three cycles.
  Background colors in panels~(b)--(d) repeat the four-stage
  partition obtained with the direction-specific $10\%$ rate threshold.}
  \label{fig:stage_resolved_output}
\end{figure*}

To assess the effect of waveform shape on mechanical output, we compare
each loaded relaxation orbit with a prescribed harmonic volume trajectory,
\begin{equation}
J_{\mathrm h}(t)=\frac{J_{\max}+J_{\min}}{2}
+\frac{\Delta J}{2}\cos\!\left(\frac{2\pi t}{P}\right),
\label{eq:harmonic_reference_trajectory}
\end{equation}
where $\Delta J=J_{\max}-J_{\min}$.  The two trajectories have the same
volume extrema, period $P$, and viscous load $\eta_L$.  The mean output
power density of the harmonic reference is
\begin{equation}
\overline q_{\mathrm h}=\frac{\pi^2\eta_L(\Delta J)^2}{2P^2}.
\label{eq:harmonic_reference_power}
\end{equation}
This reference is prescribed kinematically and is not an autonomous
solution of the thermo-swelling model.

Figure~\ref{fig:mechanical_output}(b) compares the load-pressure--volume
loops at the maximum-power load of the relaxation cycle,
$\eta_L^*=64.71\,\mathrm{MPa\,s}$.  Their enclosed areas give the extracted
work densities, which are $464$ and $217\,\mathrm{kJ\,m^{-3}}$ for the
relaxation cycle and harmonic reference, respectively.  With the common
period $P=330.65\,\mathrm{s}$, the corresponding mean power densities are
$1.403$ and $0.657\,\mathrm{kW\,m^{-3}}$.  The relaxation cycle therefore
delivers approximately $2.14$ times the work and mean power of the harmonic
reference.  At the same stroke, period, and load, this enhancement reflects
the larger mean-square volume rate produced by rapid collapse and
re-swelling.

Figure~\ref{fig:mechanical_output}(c) compares the mean mechanical output
power density of the loaded relaxation branch with that of the prescribed
harmonic reference as the viscous load is varied at $p_b=0$.  At each load,
the harmonic trajectory is matched to the volume extrema and period of
the corresponding relaxation orbit.  The relaxation waveform delivers
higher mean output power throughout the computed load range.  At the same
stroke, period, and viscous load, its rapid collapse and re-swelling
produce a larger mean-square volume rate, $\langle\dot J^2\rangle$, and
hence a higher mean output power density through
$\overline q_{\mathrm{out}}=\eta_L\langle\dot J^2\rangle$.  For both curves,
the viscous load is normalized by the relaxation load capacity,
$\eta_{\mathrm{cap}}=134.06\,\mathrm{MPa\,s}$.  The orange circle marks the
maximum-power point of the relaxation branch.

Figure~\ref{fig:mechanical_output}(d) follows the relaxation
maximum-power state under increasing opposing pressure.  Continuation gives
a finite amplitude load fold at
$p_{b,\mathrm{cap}}=58.48\,\mathrm{kPa}$.  The cycle still retains
$88.9\%$ of its zero-bias volume stroke and $79.7\%$ of its zero-bias mean
power at this boundary, so loss of oscillation under pressure is abrupt.

\subsection{Stage-resolved mechanical energy delivery}
\label{sec:stage_resolved_output}

The cycle-averaged curves in Fig.~\ref{fig:mechanical_output} do not show
when work is delivered within a relaxation cycle.  Figure~
\ref{fig:stage_resolved_output} therefore resolves the maximum-power orbit
into four stages: slow motion on the swollen branch (I), fast collapse
(II), slow motion on the collapsed branch (III), and fast re-swelling
(IV).  Figure~\ref{fig:stage_resolved_output}(a) maps these stages onto the
critical manifold, while Fig.~\ref{fig:stage_resolved_output}(b) shows the
corresponding volume trajectory over three cycles.

Figure~\ref{fig:stage_resolved_output}(c) shows that the gel pressure $p_g$
and the transmitted load pressure $p_L=\eta_L\dot J$ reverse sign between
collapse and re-swelling.  Because
$q_{\mathrm{out}}=\eta_L\dot J^2$, motion in either direction delivers
positive output power.  Figure~\ref{fig:stage_resolved_output}(d) shows that
the slow branches comprise most of the period, whereas collapse and re-swelling
produce short power bursts.  These
two fast stages occupy only $11.6\%$ of the cycle but deliver $66.0\%$ of
its work.  The canard-mediated relaxation waveform therefore localizes
mechanical energy delivery into pulses, in contrast to the smooth output of
the harmonic reference.

\subsection{Net displacement analysis for peristaltic locomotion of thermoresponsive self-oscillating gel}
\label{sec:net_displacement}

Peristaltic locomotion with net displacement is a basic motion mode of many living systems. A thermoresponsive self-oscillating hydrogel~\cite{zhao2019soft,zhang2022feedback} can undergo peristaltic locomotion with repeated long strides forward and subsequent shorter steps backward, driven by collapse (I and II in Fig.~\ref{fig:stage_resolved_output}(a)) and expansion recovery (III and IV in Fig.~\ref{fig:stage_resolved_output}(a)), respectively, of the gel. The net displacement work equation is
\begin{equation}
F_{\mathrm{net}}\,\Delta X=W_{\mathrm{collapse}}-W_{\mathrm{expansion}}.
\label{eq:net_displacement}
\end{equation}
Here $F_{\mathrm{net}}$ and $\Delta X$ denote the average net driving force
and net displacement. The quantities $W_{\mathrm{collapse}}$ and
$W_{\mathrm{expansion}}$ are the total work delivered during collapse
(stages I and II) and re-swelling (stages III and IV), respectively. Each
is obtained by integrating $q_{\mathrm{out}}$ over the corresponding
time interval and multiplying by $V_{\mathrm{ref}}$. Thus the
corresponding areas in Fig.~\ref{fig:stage_resolved_output}(d), plotted
against $t/P$, must be multiplied by $P V_{\mathrm{ref}}$. The different
collapse and re-swelling paths [Fig.~\ref{fig:stage_resolved_output}(a)]
produce distinct pressure profiles and maxima of $|p_L|$. The work
difference between the two phases supplies the net displacement work in
Eq.~\eqref{eq:net_displacement}.

\section{Conclusion}\label{sec:conclusion}

In this work, we develop a two-variable, spatially lumped dynamical model of a thermoresponsive self-oscillating hydrogel, in which thermally driven contraction is followed by swelling induced by heat dissipation. A key feature of the model is the use of a Hill function to implement a feedback switch that modulates heat flow in response to deformation. Through linear stability analysis, nonlinear bifurcation analysis, and singular perturbation theory of the underlying geometric manifolds, we identify a continuous transition from a supercritical Hopf bifurcation to relaxation oscillations mediated by a canard explosion.

Furthermore, by analyzing the deformation velocity--driving pressure relationship within the thermoresponsive self-oscillating actuator model, we quantify the mechanical power generated during contraction and swelling, as well as the net work produced over one oscillatory cycle. These results provide a mechanistic explanation for the reciprocal, wormlike locomotion of a thermoresponsive self-oscillating actuator. This work establishes a theoretical foundation for the further development of thermally driven self-oscillating soft actuators and their functional applications.

\begin{acknowledgments}
Q.G. acknowledges financial support from the National Natural Science Foundation of China (grant no. 22120102001) and (grant no. 22372122), and the National Key Research and Development Program of China (Grant No.2026YFE0101500).
L.R. acknowledges financial support from the National Natural Science Foundation of China (Grant No. 22372122).
The authors acknowledge the interdisciplinary science and engineering research environment provided by the China University of Mining and Technology.
\end{acknowledgments}

\section*{Data Availability}
The data that support the findings of this article are not publicly available.
The data are available from the authors upon reasonable request.

\appendix

\section{Constitutive basis and lumped-model reduction}
\label{app:model_derivation}

\subsection{Free-swelling equilibrium}
Within a continuum description of gel swelling~\cite{onuki1993theory,barriere2003kinetics,doi2009gel}, the local free-energy density is expressed in terms of the polymer volume fraction $\phi(\mathbf{x},t)$, the left Cauchy--Green tensor $\hat{\mathbf B}$, and the temperature $T$. We combine the rubber-elastic contribution used in Ref.~\cite{he2012synthetic} with the composition-dependent PNIPAAm--water mixing free energy introduced in Sec.~\ref{sec:model}. With $I_1=\mathrm{tr}\,\hat{\mathbf B}$ and $I_3=\det\hat{\mathbf B}$, conservation of polymer volume gives
\begin{equation}
\phi=\phi_0 I_3^{-1/2}.
\end{equation}
The dimensionless reference-volume free-energy density combines rubber elasticity and Flory--Huggins mixing~\cite{flory1943statistical,quesada2011gel},
\begin{equation}
U=U_{\mathrm{el}}+U_{\mathrm{FH}},
\end{equation}
where
\begin{align}
U_{\mathrm{el}} &= \tfrac{1}{2}c_0 v_0\!\left[I_1-3-\ln\!\left(I_3^{1/2}\right)\right], \label{eq:Uel}\\
U_{\mathrm{FH}} &= I_3^{1/2}\!\left[(1-\phi)\ln(1-\phi)+\chi(\phi,T)\phi(1-\phi)\right]. \label{eq:UFH}
\end{align}
Here $\chi(\phi,T)$ is the interaction parameter defined in Eq.~\eqref{eq:chi}. Because it depends on $\phi$, its concentration derivative must be retained when the free energy is differentiated. The corresponding dimensionless Cauchy stress is
\begin{equation}
\hat{\boldsymbol\sigma}
=-\left[\pi_{\mathrm{osm}}(\phi,T)+c_0v_0\frac{\phi}{2\phi_0}\right]\hat{\mathbf I}
+c_0v_0\frac{\phi}{\phi_0}\hat{\mathbf B}.
\end{equation}
For homogeneous isotropic free swelling, $\hat{\mathbf B}=\lambda^2\hat{\mathbf I}$ and $\lambda=(\phi_0/\phi)^{1/3}$. Taking the bath pressure as the zero of gauge stress, zero applied traction requires the total Cauchy stress to satisfy $\hat{\boldsymbol\sigma}=0$; the elastic and mixing contributions remain nonzero and balance one another. Substitution gives $\Pi_{\mathrm{el}}=\pi_{\mathrm{osm}}$ with the contributions in Eqs.~\eqref{eq:Pi_el} and \eqref{eq:pi_osm}. The concentration derivative above is the origin of the additional $\chi_1$ terms in Eq.~\eqref{eq:pi_osm}. Thus the zeros of Eq.~\eqref{eq:0d_final_phi} reproduce the free-swelling equilibria associated with this free-energy density.

\subsection{Equilibrium calibration and uncertainty}
\label{app:equilibrium_calibration}

The working material point in Fig.~\ref{fig:equilibrium_combined}(b) is
calibrated to the lowest crosslink series ($C=1.23\%$) reproduced in
Fig.~4 of Oh~\cite{oh1998crosslink}.  Marker
centers were extracted from the PDF vector paths using the vector tick
coordinates rather than manual clicking; the data set identity and
$\phi_0=0.048$ were checked against the original paper.  With $V_0$
denoting the dry polymer volume, the reported volume ratios give $\phi=V_0/V$.

\begin{figure}
  \centering
  \includegraphics[width=\columnwidth]{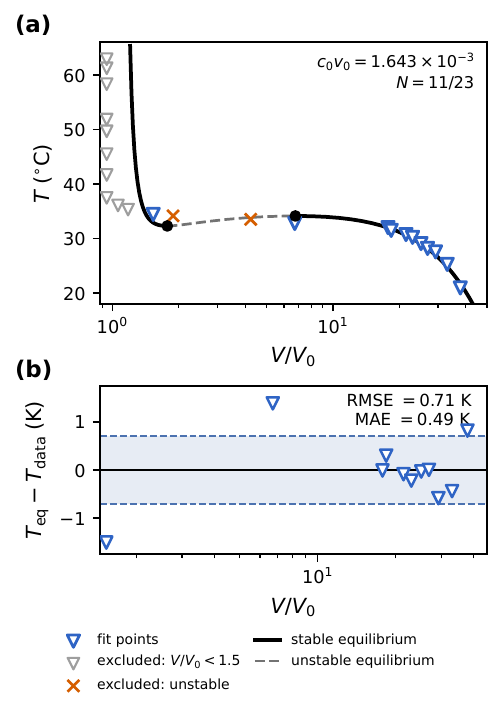}
  \caption{Equilibrium calibration used for the working material point.
  (a)~$C=1.23\%$ data and fitted manifold.  Blue triangles enter the
  fit; gray triangles fail $V/V_0\geq1.5$, and orange crosses fail the local
  stability test.  Solid and dashed curves are stable and unstable
  equilibria, respectively, and circles mark the folds.
  (b)~Temperature residuals for the selected points; the band denotes $\pm$RMSE.}
  \label{fig:equilibrium_calibration}
\end{figure}

With the four interaction coefficients and $\phi_0$ fixed, only $c_0v_0$
was fitted.  With $\phi_i=V_0/V_i$ for each observation, we minimized
$\sum_i2s_T^2[\sqrt{1+(r_i/s_T)^2}-1]$, where
$r_i=T_{\rm eq}(\phi_i;c_0v_0)-T_i$ and $s_T=1\,\mathrm{K}$.  We excluded
$V_i/V_0<1.5$, for which marker uncertainty is largest and the interaction-law
parameterization is least applicable, and required
$g_\phi(\phi_i,T_i)<0$ so that stable observations were not fitted to the
repelling branch.  This retains 11 of 23 points; ten fail the volume cutoff
and two fail the stability test.  These are model scope rather than residual
criteria, and all observations remain visible in
Fig.~\ref{fig:equilibrium_calibration}(a).

The fit gives $c_0v_0=1.6425\times10^{-3}$, with a conditional 95\% interval
$[1.5623,1.7187]\times10^{-3}$ from 1000 nonparametric bootstrap resamples.
For the selected points, the RMSE and mean absolute error are $0.709$ and
$0.490\,\mathrm{K}$.  The same one-parameter procedure gives RMSE values of
$0.906$, $1.318$, and $0.921\,\mathrm{K}$ for the other two series and
the Suzuki series, respectively, as cross-material checks.  The interval is
conditional on the fixed interaction law, $\phi_0$, digitization, and
selection rule; it neither includes their systematic uncertainty nor
calibrates the dynamic or feedback parameters.

\subsection{Lumped swelling kinetics and thermal balance}
\label{app:lumped_kinetics}
The overdamped two-fluid description gives the polymer-network velocity
as~\cite{he2012synthetic,barriere2003kinetics}
\begin{equation}
\mathbf{v}^{(p)}
=
\Lambda\left(\frac{\phi}{\phi_0}\right)^{-3/2}(1-\phi)\nabla\cdot\hat{\boldsymbol\sigma},
\end{equation}
where $\Lambda$ is a mobility prefactor. Polymer conservation gives
\begin{equation}
\frac{\partial\phi}{\partial t}
=
-\nabla\cdot\left(\phi\mathbf{v}^{(p)}\right).
\end{equation}
For strictly uniform fields, $\nabla\cdot\hat{\boldsymbol\sigma}=0$, so these equations do not directly yield nontrivial zero-dimensional dynamics. We therefore use Eq.~\eqref{eq:0d_final_phi} as a phenomenological overdamped law for the spatially averaged swelling coordinate. It preserves the free-swelling equilibrium manifold, while $\tau_\phi$ absorbs mobility, specimen size and geometry, and boundary solvent transport. Spatial transients remain outside the scope of the model.

For a spatially uniform temperature, constant effective heat capacity $C_{\mathrm{th}}$ and thermal conductance $H_{\mathrm{th}}$ give
\begin{equation}
C_{\mathrm{th}}\frac{dT}{dt}=Q_0 f(\phi)-H_{\mathrm{th}}(T-T_0).
\label{eq:energy_balance_dim}
\end{equation}
Here $Q_0$ is the plateau heating power. Reference-state estimates are $C_{\mathrm{th}}\simeq\rho_m c_pV_{\mathrm{ref}}$ and $H_{\mathrm{th}}\simeq h_{\mathrm{th}}A_{\mathrm{ref}}$, where $\rho_m$, $c_p$, $h_{\mathrm{th}}$, $V_{\mathrm{ref}}$, and $A_{\mathrm{ref}}$ denote density, specific heat, heat transfer coefficient, reference volume, and reference area. Dividing by $C_{\mathrm{th}}$ gives Eq.~\eqref{eq:0d_final_T}, with $\gamma=Q_0/C_{\mathrm{th}}$ and $k=H_{\mathrm{th}}/C_{\mathrm{th}}$. The equilibrium calibration described in Sec.~\ref{sec:model} determines neither $\gamma$ nor $k$; possible state dependence of the effective thermal coefficients is neglected.

Finally, the shifted Hill gate in Eq.~\eqref{eq:gate} satisfies $f=1$ for $\phi\le\phi_g$ and decreases as the gel becomes denser. It is $C^1$ at $\phi_g$ for $n\ge2$ and $C^2$ there for the chosen value $n=3$. All Hopf equilibria analyzed in the main text lie strictly above $\phi_g$, where the gate is smooth and the third derivatives required for the first Lyapunov coefficient are well defined.

\begin{figure}[tb]
  \centering
  \includegraphics[width=\columnwidth]{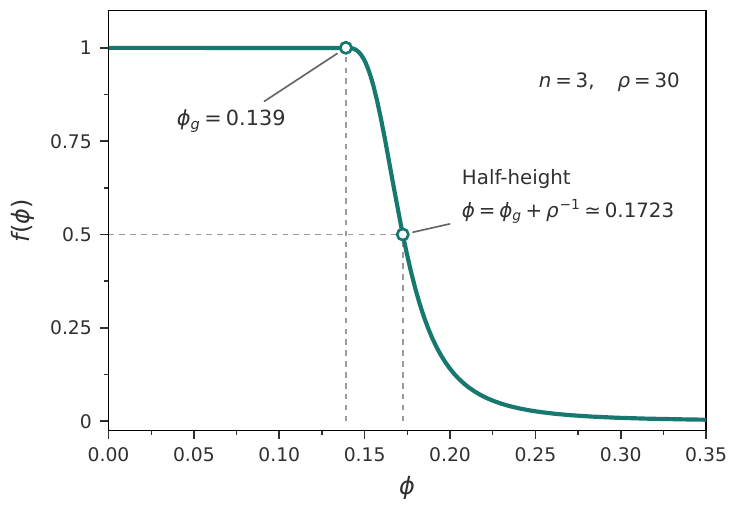}
  \caption{Shifted Hill heating gate in Eq.~\eqref{eq:gate} for
  $\phi_g=0.139$, $\rho=30$, and $n=3$.
  Open circles mark the plateau boundary $\phi=\phi_g$ and the half-height
  point $\phi=\phi_g+\rho^{-1}\simeq0.1723$, where $f=1/2$.}
  \label{fig:heating_gate}
\end{figure}

\section{Equilibrium stability formulas}
\label{app:equilibrium_stability}

\subsection{Closed-form derivatives}
The slope $g_\phi(\phi,T)=\partial g/\partial\phi$ controls the stability of
the fast mechanical subsystem at fixed temperature; stability of the full
coupled fixed point follows from the trace and determinant in
Appendix~\ref{app:linear_stability}~\cite{flory1943statistical,quesada2011gel,lopez2017flory}:
\begin{align}
g_\phi &= c_0v_0\!\left[\frac{1}{3\phi_0}\!\left(\frac{\phi}{\phi_0}\right)^{\!-2/3}\!-\frac{1}{2\phi_0}\right] \nonumber\\
       &\quad +\left[1-\frac{1}{1-\phi}+2(\chi_0-\chi_1)\phi
+6\chi_1\phi^2\right],
\end{align}
where $\chi_0$ and $\chi_1$ are evaluated at $T_K$.
Bistability vanishes at the cusp point defined by $g=g_\phi=g_{\phi\phi}=0$, where
\begin{align}
g_{\phi\phi} &= c_0v_0\!\left[-\frac{2}{9\phi_0^2}\!\left(\frac{\phi}{\phi_0}\right)^{\!-5/3}\right] \nonumber\\
             &\quad +\left[-\frac{1}{(1-\phi)^2}+2(\chi_0-\chi_1)
+12\chi_1\phi\right].
\end{align}

\subsection{Fold curves and material map with fixed interaction parameters}
Because $\chi_0$ and $\chi_1$ are affine in $T_K$, define
\begin{align}
E(\phi)&=\left(\frac{\phi}{\phi_0}\right)^{1/3}-\frac{\phi}{2\phi_0},\nonumber\\
R(\phi)&=\phi+\ln(1-\phi)+(A_0-A_1)\phi^2+2A_1\phi^3,\nonumber\\
D(\phi)&=(B_0-B_1)\phi^2+2B_1\phi^3.
\label{eq:phase_map_affine}
\end{align}
The equilibrium condition becomes
\begin{align}
N(\phi)&=c_0v_0E(\phi)+R(\phi),\nonumber\\
g(\phi,T_K)&=N(\phi)+T_KD(\phi).
\label{eq:fold_affine_temperature}
\end{align}
Consequently, the equilibrium manifold has the explicit representation
\begin{align}
T_{{\rm eq},K}(\phi)=-\frac{N(\phi)}{D(\phi)},
\label{eq:equilibrium_temperature_explicit}
\end{align}
with $T_{\rm eq}(\phi)=T_{{\rm eq},K}(\phi)-273.15\,^\circ\mathrm C$ in
the Celsius convention used in the main text. Its folds satisfy
\begin{equation}
N'(\phi)D(\phi)-N(\phi)D'(\phi)=0.
\label{eq:fold_temperature_condition}
\end{equation}
For a fixed elastic coefficient, the two solutions of
Eq.~\eqref{eq:fold_temperature_condition} give the lower and upper fold
temperatures plotted in Fig.~\ref{fig:equilibrium_combined}(c). The same
functions also give a direct parameterization of these curves with all
four mixing coefficients fixed.
At fixed $\phi$, solving $g=g_\phi=0$ for $(c_0v_0,T_K)$ gives
\begin{align}
\Delta(\phi)&=E(\phi)D'(\phi)-E'(\phi)D(\phi),\nonumber\\
c_0v_0(\phi)&=\frac{D(\phi)R'(\phi)-D'(\phi)R(\phi)}{\Delta(\phi)},\nonumber\\
T_K(\phi)&=\frac{E'(\phi)R(\phi)-E(\phi)R'(\phi)}{\Delta(\phi)}.
\label{eq:phase_map_fold_surface}
\end{align}
Sweeping $\phi$ traces both fold branches in the $(c_0v_0,T)$ plane without treating the state-dependent $\chi_1(T)$ as an independent coordinate. Their common endpoint additionally satisfies $g_{\phi\phi}=0$ and occurs at $c_0v_0=2.4801\times10^{-2}$, $T=38.442\,^\circ$C, and $\phi=0.3681$.

\section{Hopf normal form and Lyapunov coefficient}
\label{app:lyapunov}

\subsection{Jacobian entries and Hopf conditions}
\label{app:linear_stability}

The Jacobian of Eqs.~\eqref{eq:0d_final_phi} and
\eqref{eq:0d_final_T} at a fixed point $(\phi^\ast,T^\ast)$ in physical time is
\begin{equation}
\mathbf J_\ast=\begin{pmatrix}
g_\phi/\tau_\phi & -\pi_T/\tau_\phi\\
\gamma f'(\phi) & -k
\end{pmatrix}_\ast.
\label{eq:J_entries}
\end{equation}
The trace and determinant read
\begin{align}
\mathrm{tr}\,\mathbf J_\ast &= g_\phi/\tau_\phi-k,\label{eq:traceJ}\\
\det\mathbf J_\ast        &= -(k/\tau_\phi)g_\phi+(\gamma/\tau_\phi)f'(\phi^\ast)\pi_T(\phi^\ast,T^\ast).\label{eq:detJ}
\end{align}
The Hopf conditions $\mathrm{tr}\,\mathbf J_\ast=0$ with $\det\mathbf J_\ast>0$
[Eq.~\eqref{eq:hopf_conditions}] follow by direct substitution, with Hopf
frequency $\omega_H=\sqrt{\det\mathbf J_\ast}$.
To obtain the fold expansion used in
Eq.~\eqref{eq:hopf_fold_prediction}, write the equilibrium manifold as
$T=T_{\rm eq}(\phi)$ and
$\alpha_{\rm eq}(\phi)=[T_{\rm eq}(\phi)-T_0]/f(\phi)$.
Writing the swollen fold as $F_{\rm sw}=(\phi_F^{\rm sw},T_+)$ gives
$\alpha_F=\alpha_{\rm eq}(\phi_F^{\rm sw})
=(T_+-T_0)/f(\phi_F^{\rm sw})=9.31615\,^\circ\mathrm C$.
Expanding $g_\phi=\varepsilon$ about $F_{\rm sw}$ gives
\begin{align}
\phi_H-\phi_F^{\rm sw}
&=\frac{\varepsilon}{g_{\phi\phi,F}}+O(\varepsilon^2),\nonumber\\
\alpha_H-\alpha_F
&=\frac{\alpha'_{\rm eq,F}}{g_{\phi\phi,F}}\varepsilon
+O(\varepsilon^2).
\end{align}
For the data-constrained constitutive law,
$g_{\phi\phi,F}=0.237768$ and
$\alpha'_{\rm eq,F}=60.0810\,^\circ\mathrm C$, giving the coefficient
$252.688\,^\circ\mathrm C$ quoted in the main text.

For the affine interaction law in Eq.~\eqref{eq:chi},
\begin{align}
\chi_0'(T)&=B_0,& \chi_1'(T)&=B_1,\\
\chi_0''(T)&=\chi_1''(T)=0,&
\chi_0'''(T)&=\chi_1'''(T)=0,
\end{align}
where derivatives with respect to degrees Celsius and kelvin increments are identical. Hence
\begin{equation}
\pi_T=-\left[(B_0-B_1)\phi^2+2B_1\phi^3\right],
\qquad \pi_{TT}=0,
\end{equation}
while the mixed $\phi$--$T$ derivatives remain nonzero and must be retained in the multilinear forms below. The gate derivative follows from Eq.~\eqref{eq:gate}: with $x\equiv\max(\phi-\phi_g,0)$,
\begin{equation}
f'(\phi)=\begin{cases}
0 & \phi\le\phi_g,\\[2pt]
\displaystyle-\,\frac{n\rho(\rho x)^{n-1}}{[1+(\rho x)^n]^2} & \phi>\phi_g.
\end{cases}
\end{equation}
For $n=3$ this reads
$f'(\phi)=-3\rho^3 x^2/[1+(\rho x)^3]^2$ for $\phi>\phi_g$.

\subsection{Lyapunov coefficient formulas}
For the physical-time vector field $\mathbf F$ of
Eqs.~\eqref{eq:0d_final_phi} and \eqref{eq:0d_final_T}, with
$\mathbf x=(\phi,T)^\mathsf{T}$, write $\mathbf J=\mathbf J_\ast$ at the
Hopf equilibrium $\mathbf x^\ast$ and expand
\begin{multline}
  \mathbf{F}(\mathbf{x}^\ast+\mathbf{u})
  = \mathbf J\mathbf{u}
  + \tfrac{1}{2}B(\mathbf{u},\mathbf{u})
  + \tfrac{1}{6}C(\mathbf{u},\mathbf{u},\mathbf{u}) \\
  + O(\|\mathbf{u}\|^4),
\end{multline}
where $B$ and $C$ are the symmetric multilinear forms of the second and third derivatives of $\mathbf{F}$ at $\mathbf{x}^\ast$:
\begin{align}
  B_k(\mathbf{u},\mathbf{v})
  &= \sum_{i,j}\frac{\partial^2 F_k}{\partial x_i\partial x_j}\bigg|_{\ast} u_i v_j, \nonumber\\
  C_k(\mathbf{u},\mathbf{v},\mathbf{w})
  &= \sum_{i,j,l}\frac{\partial^3 F_k}{\partial x_i\partial x_j\partial x_l}\bigg|_{\ast} u_i v_j w_l.
  \label{eq:multilinear}
\end{align}
Let $q$ be the right eigenvector of $\mathbf J$ with eigenvalue $i\omega_H$ and $p$ the adjoint eigenvector satisfying $\langle p,q\rangle=1$, where $\langle\cdot,\cdot\rangle$ denotes the standard Hermitian inner product. Define the auxiliary vectors
\begin{align}
  h_{20} &= (2i\omega_H I - \mathbf J)^{-1}\,B(q,q), \nonumber\\
  h_{11} &= -\mathbf J^{-1}\,\operatorname{Re}\bigl[B(q,\bar{q})\bigr].
  \label{eq:h20_h11}
\end{align}
The first Lyapunov coefficient is then~\cite{kuznetsov2004elements}
\begin{multline}
  l_1 = \frac{1}{2\omega_H}\operatorname{Re}\bigl[
    \langle p, C(q,q,\bar q)\rangle
    + 2\langle p, B(q,h_{11})\rangle \\
    + \langle p, B(\bar q,h_{20})\rangle
  \bigr].
  \label{eq:L1_formula}
\end{multline}

\subsection{Hopf criticality on the representative slice}
For $\tau_\phi=0.1\,\mathrm{s}$,
$k=5\times10^{-3}\,\mathrm{s}^{-1}$, and hence
$\varepsilon=5\times10^{-4}$, solving the equilibrium and Hopf conditions
gives $\alpha_H=9.47450755\,^\circ\mathrm C$ at
$(\phi_H,T_H)=(0.150119,34.1354\,^\circ\mathrm C)$.  The Hopf frequency is
$0.0916606\,\mathrm{rad\,s^{-1}}$, corresponding to a period of
$68.5484\,\mathrm{s}$.  Evaluation of
Eq.~\eqref{eq:L1_formula} from the exact derivatives of the data-constrained
constitutive law and the smooth side of the Hill gate gives
$l_1=-43.4168$ in the adopted coordinates and time normalization.  The
magnitude changes under state and time rescaling, whereas the negative sign
is the invariant information used in the main text: the Hopf point is
supercritical.

\section{Numerical methods}
\label{app:numerical_methods}

\begin{figure*}
  \centering
  \includegraphics[width=\linewidth]{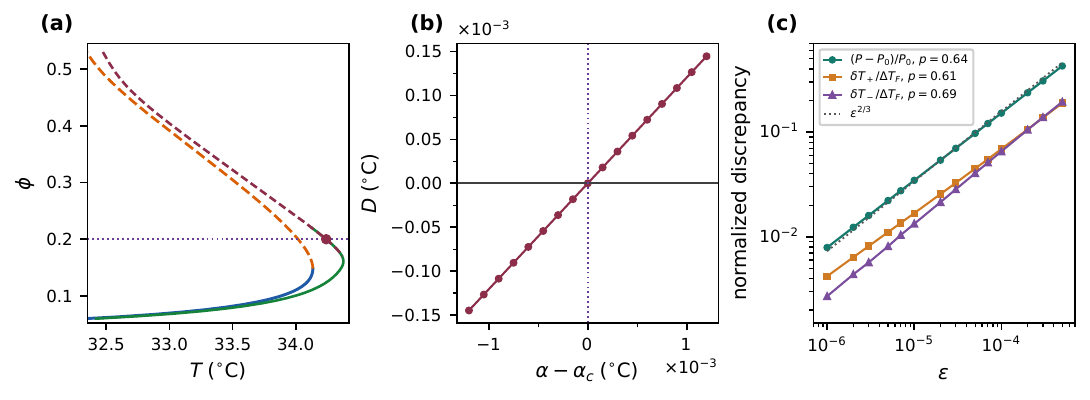}
  \caption{Numerical validation of the fast--slow constructions.
  Panels~(a) and (b) show finite-$\varepsilon$ slow-manifold matching at
  $\varepsilon=5\times10^{-4}$.  (a)~The attracting sheets of the critical
  manifold $\mathcal M_0$ are blue and the repelling sheet is orange dashed.
  The finite-$\varepsilon$ attracting slow manifold (green) is integrated
  forward from the polymer-poor sheet, whereas the repelling slow manifold
  (burgundy dashed) is integrated backward from the middle sheet.  The purple dotted line is the transverse
  section $\Sigma:\phi=0.2$ and the dot marks their matched intersection at
  $\alpha=\alpha_c$.  (b)~Signed temperature separation $D$ (attracting minus repelling) on
  $\Sigma$ versus distance from $\alpha_c$.  Its zero crossing
  selects the maximal canard independently of any periodic-orbit amplitude
  threshold; the axis multipliers expose the narrow numerical scale of the
  matching.
  (c)~Finite-rate convergence at $\alpha=14\,^\circ\mathrm C$.  The relative
  period excess and the two normalized fold overshoots approach zero with
  fitted exponents close to the $\varepsilon^{2/3}$ reference scaling, where
  $\Delta T_F=T_+-T_-$,
  $\delta T_+=T_{\max}-T_+$, and
  $\delta T_-=T_--T_{\min}$.}
  \label{fig:maximal_canard_matching}
\end{figure*}

\subsection{Direct integration and validation}
Direct integration is used only to seed and independently validate the
continued periodic orbits, not to locate either the Hopf or maximal-canard
locus.  Stable-cycle seeds are computed in physical time with the SciPy
\texttt{solve\_ivp} DOP853 method, while representative small- and
large-amplitude cycles and the right edge of
Fig.~\ref{fig:hopf_canard_organization}(c) are recomputed in slow time with
the stiff Radau method.  Periods are evaluated after transients from
successive maxima.  The largest cycle-closure residual among these checks is
$1.1\times10^{-8}$.

\subsection{Signed first Lyapunov coefficient}
Hopf points are obtained by solving the equilibrium condition together with
$\mathrm{tr}\,\mathbf J=0$ and retaining solutions with $\det\mathbf J>0$.  The signed
$l_1$ is evaluated from Eq.~\eqref{eq:L1_formula} using both exact derivative
tensors and centered finite differences over a sequence of step
refinements.  On the $\varepsilon=5\times10^{-4}$ slice, every refinement
reproduces the negative sign of the exact result, with maximum relative
difference $5.3\times10^{-5}$.

\subsection{Periodic orbit continuation and maximal canard matching}

\begingroup
\looseness=-1
Periodic orbits underlying
Figs.~\ref{fig:hopf_canard_organization}(a)--(c) and
\ref{fig:canard_representative_states} are represented on a normalized
phase interval and solved with the fourth-order adaptive collocation scheme
in SciPy \texttt{solve\_bvp}.  The period is an unknown and
$\dot\phi(0)=0$ fixes the time phase.
Near the almost vertical canard segment, $\phi_{\max}$ rather than $\alpha$
is used as the continuation coordinate.  Fully developed post-canard
cycles are continued at fixed $\alpha$ by Poincar\'e-return shooting.  For a \mbox{planar} autonomous flow the
nontrivial Floquet multiplier is
\begin{equation}
\mu_\perp=\exp\left[\int_0^P\nabla\!\cdot\mathbf F\,dt\right],
\label{eq:planar_floquet}
\end{equation}
which provides an independent stability test.  Across the periodic orbits
used in Fig.~\ref{fig:hopf_canard_organization}(c), the largest collocation
RMS residual is $1.0\times10^{-6}$, and no multiplier-one crossing is
detected away from the Hopf endpoint.\par
\endgroup

The finite $\varepsilon$ maximal canard is located directly from slow
manifolds rather than from an amplitude threshold.  Away from the folds,
the first-order graph
$T_\varepsilon(\phi)=T_{\rm eq}(\phi)
-\varepsilon h[\phi,T_{\rm eq}(\phi);\alpha]/
g_\phi[\phi,T_{\rm eq}(\phi)]+O(\varepsilon^2)$ initializes forward and
backward integration of the attracting and repelling slow manifolds,
respectively.  The zero of their signed temperature separation on the
transverse section $\Sigma:\phi=0.2$ defines $\alpha_c$.  At
$\varepsilon=5\times10^{-4}$ this gives
$\alpha_c=12.06036\,^\circ\mathrm C$.  Varying the section over
$0.18\leq\phi\leq0.22$ and changing the initialization points displaces
$\alpha_c$ by at most $2.10\times10^{-6}\,^\circ\mathrm C$; periodic
collocation independently confirms the matched value.  The construction is
shown in Fig.~\ref{fig:maximal_canard_matching}.

For Fig.~\ref{fig:hopf_canard_organization}(c), the Hopf and
maximal-canard calculations are repeated over
$2\times10^{-5}\leq\varepsilon\leq7\times10^{-4}$.  The Hopf curve follows
from the equilibrium, zero trace, and positive determinant conditions,
whereas slow manifold matching gives the maximal canard locus, which is
joined to the analytical endpoint $(\alpha_F,0)$.  Every sampled Hopf point
is supercritical, and the continued periodic branch is stable.

\subsection{Robustness to the feedback-gate closure}
\label{app:gate_robustness}

At $\varepsilon=5\times10^{-4}$, we repeated the Hopf,
first Lyapunov coefficient, slow manifold matching, and post-canard
calculations for one-at-a-time variations
$\phi_g=0.133,0.139,0.145$, $\rho=20,30,40$, and $n=2,3,4$, and for
logistic and exponential-plateau gates matched at the baseline half-height
and slope.  All nine closures retain a supercritical Hopf point, a maximal
canard, and a stable post-canard relaxation cycle, with
$9.17\lesssim\alpha_H\lesssim10.37\,^\circ\mathrm C$ and
$10.66\lesssim\alpha_c\lesssim13.15\,^\circ\mathrm C$.  The alternatives
contain short unstable segments inside the canard interval, but recover
stable large-amplitude branches.  Hence the sequence is structurally
robust, whereas its thresholds and detailed Floquet stability are closure
dependent.  Complete definitions, branches, validation tables, and the
comparison figure are supplied with the reproducibility archive.

\subsection{Finite-rate convergence to the singular limit}
\label{app:singular_convergence}

The finite-rate correction to Fig.~\ref{fig:singular_limit_prediction} is
tested at fixed $\alpha=14\,^\circ\mathrm C$ and
$k=5\times10^{-3}\,\mathrm{s}^{-1}$ over
$10^{-6}\leq\varepsilon\leq5\times10^{-4}$ by varying $\tau_\phi$ alone.
Stiff slow-time integration with Radau and log--log fits over
$\varepsilon\leq10^{-4}$ give exponents $0.641$, $0.606$, and $0.693$ for
the relative period excess and the normalized upper and lower fold
overshoots, respectively [Fig.~\ref{fig:maximal_canard_matching}(c)].
These values are consistent with the generic
$O(\varepsilon^{2/3})$ fold-layer scaling.  By
$\varepsilon=10^{-6}$ the period differs from the singular quadrature by
less than $1\%$, while the relative amplitude difference is below
$3\times10^{-4}$.

\FloatBarrier
\section{Mechanical output continuation and validation}
\label{app:mechanical_output_numerics}

The load transformation used in Sec.~\ref{sec:load_model} follows directly
from $J=\phi_0/\phi$ and Eq.~\eqref{eq:0d_final_phi}:
\begin{equation}
\dot J=-\frac{\phi_0}{\phi^2}\dot\phi
=\frac{\phi_0}{\tau_\phi p_T(T)\phi^2}p_g.
\label{eq:load_transformation}
\end{equation}
This identity gives the state-dependent internal viscosity specified below
Eq.~\eqref{eq:gel_output_pressure} without an additional constitutive
assumption.  Multiplying Eq.~\eqref{eq:loaded_gel_balance} by $\dot J$ and
integrating over one period gives Eq.~\eqref{eq:mechanical_energy_identity};
the bias term vanishes because
$p_b\int_0^P\dot J\,dt=p_b\oint dJ=0$.  A positive $p_b$ therefore opposes
increasing $J$ but performs no net work over a closed reciprocal cycle.

Loaded periodic orbits are seeded by Radau integration and then solved at
fixed $(\alpha,k,\eta_L,p_b)$ by fourth-order adaptive collocation on a
normalized phase interval.  The period is an unknown, periodicity supplies
two boundary conditions, and
$\dot\phi(0)=0$ fixes the time phase at a maximum of $\phi$.  The no-load
solutions reproduce the periods and amplitudes of the independently
continued autonomous branches with relative errors below
$2.8\times10^{-8}$.

For a planar autonomous system, the nontrivial Floquet multiplier is
Eq.~\eqref{eq:planar_floquet}.  A fold of loaded cycles is located by
augmenting the boundary-value problem with
\begin{equation}
\dot z=\nabla\!\cdot\mathbf F_L(\phi,T),\qquad
z(0)=z(P)=0,
\end{equation}
where $\mathbf F_L$ is the loaded vector field defined by
Eq.~\eqref{eq:loaded_gel_balance} and the heat balance.  The load
coefficient or the pressure bias and the period are solved simultaneously.
The auxiliary Floquet constraint verifies the two folds reported in
Sec.~\ref{sec:waveform_output}.  Work metrics are evaluated by quadrature on
the periodic solutions, and maximum power loads are bracketed on the
continued branch before refinement as boundary value problems.  Across the
load and pressure bias calculations, the energy identity and periodic
closure are satisfied to numerical precision, and the largest collocation
RMS residual is $2.0\times10^{-6}$.

\begingroup
\looseness=-1
For the stage-resolved diagnostic in
Fig.~\ref{fig:stage_resolved_output}, $\dot J$ is recovered directly from
$p_L/\eta_L$ on the loaded collocation orbit and
$\dot\phi=-(\phi^2/\phi_0)\dot J$; no finite-difference rate is used.
Fast collapse and fast re-swelling are the circular connected components
around the positive and negative extrema of $\dot\phi$ at $10\%$ of the
corresponding directional peak.  Varying this threshold from $5\%$ to
$15\%$ leaves the conclusion unchanged: the two fast components account for
$58.9\%$--$76.3\%$ of the extracted work.  The threshold is used only for
the stage-resolved energy budget, not to locate any dynamical boundary.
The archive contains tolerances, mesh limits, and convergence records.\par
\endgroup

\bibliographystyle{apsrev4-2-title}
\bibliography{references}

\end{document}